# Automated Extraction of Unstructured Post-SBRT Toxicity Data from Radiology Reports Using Large Language Models

Justin Pijanowski[1], Yakout Mezgueldi[1], Alan Lee[1], Drew Moghanaki[1], Ricky R. Savjani[1], James Lamb[1]

Department of Radiation Oncology, University of California, Los Angeles[1]

**Abstract**

We evaluated the viability of using a Large Language Model (LLM) to extract patient-specific specific toxicity and progression outcomes from unstructured radiology reports. We retrospectively extracted 160 follow-up CT and PET/CT electronic medical record notes for patients treated with lung stereotactic body radiotherapy (SBRT) at our institution from January 2017 through December 2023. Using the Llama 3.3-70-B-Instruct LLM, we engineered prompts to extract four clinical endpoints from each radiology report: locoregional progression, distant progression, radiation-induced fibrosis, and radiation-induced rib fractures. Progression endpoints were classified as yes, no, or maybe, while fibrosis and rib fractures were binary (yes or no). Ground truth labels were defined using two-grader consensus for the 60-note training set, used for prompt development, and a three-grader majority vote for the 100-note test set. LLM performance was evaluated using sensitivity, specificity, and accuracy. As detailed by our evaluation metrics, the strong performance of our methods demonstrates the viability of using prompt-engineered LLMs to extract radiation-toxicities and progression classification from radiology reports.

# 1 Introduction

Natural language processing (NLP) is a domain within computer science focused on creating algorithms that enable computers to understand and analyze human language[1]. Named entity recognition (NER) is a NLP technique used to extract discrete pieces of information from bodies of text[2]. Traditional NLP algorithms, defined as algorithms developed before the advent of the transformer architecture[3], relied on rule-based techniques, probabilistic methods, and manual feature engineering to extract textual feature representations[4,5]. Frequent applications included information retrieval, text generation, sentiment analysis, and text analysis[6,7]. These traditional NLP algorithms were computationally lightweight, easy to understand, and implemented with ease. As NLP methods advanced, the extracted textual feature representations were incorporated into machine learning classifiers for sophisticated text classification and analysis. Deep neural network based NLP models, such as those utilizing recurrent and convolutional networks, made significant improvements in capturing long range textual dependencies and generating richer semantic understanding[8,9]. Early NER techniques were based on these traditional NLP algorithms, however, the development of the transformer architecture significantly advanced NER performance and language comprehension[4,5]. Large Language Models (LLMs) are deep artificial intelligence networks built upon the transformers architecture and pretrained on extensive amounts of textual data[10]. LLMs have revolutionized NLP and performed tasks such as translation between languages, assistance with literature reviews, and mining data from databases[11–15].

Large, unstructured datasets contain valuable information that is difficult to extract and analyze using traditional structured data tools. LLMs resolve these difficulties by extracting targeted textual information and transforming the unstructured datasets into structured and predictable formats for efficient data analysis[16]. A patient's electronic health record (EHR) contains their treatment and medical history, including lab results, physician reports and treatment summaries, progress notes, and more. Centralizing medical data from multiple sources into an EHR provides both the patients and healthcare workers with easy access to health records and makes it possible to actively track and understand one's own medical history. EHR data is stored in structured and unstructured formats. Structured data, which includes patient demographics, laboratory results, and medical codes, are stringently organized to allow for accessible querying and analysis. Unstructured data, such as radiology report note text, progress notes, and medical imaging, is challenging to extract but contains critical information for patient analysis [17].

LLMs have been used in medicine for numerous tasks, including responding to medical questions[18,19], supporting research[20,21], assisting in medical education[22], and increasing accessibility to medical knowledge[23,24]. A prominent application is the mining of information from unstructured data in the EHR. While traditional NLP algorithms have been used for clinical NER[25,26], recent research has adopted LLMs for these same tasks.

Recent studies have used LLMs for clinical documentation support, including automated clinical note generation and real-time emergency medicine documentation[27–30]. Additionally, LLMs were applied to extract clinically relevant concepts from clinical free-text notes, including social determinants of health[31,32], disease progression information from admission records using prompt engineering[33], and oncologic information, such as cancer diagnosis, tumor staging, histology type, and treatment history using zero-shot and prompt based approaches[34,35]. Other work used prompt engineering to detect acute myocardial infarction, diabetes, and hypertension from healthcare notes[36], and the extraction of outcomes and response assessments from radiology and pathology reports[37,38]. Furthermore, research demonstrated that LLMs can successfully extract structured information such as the helmet status of individuals with micro-mobility related injuries[39], numeric variables from organ procurement notes[40], and cognitive exam dates and scores[41].

The use of LLMs in radiation oncology is rapidly expanding to the domains of treatment planning, workflow automation, and clinical communication. Recent work integrated an LLM extraction tool from clinical notes to supplement their deep learning framework for automatic contouring of a three-dimensional clinical target volume[42]. Radiation oncology specific models, such as RadOnc-GPT, can utilize patient diagnostic descriptions to support treatment planning, including using patient details to generate initial plans, determining the optimal treatment modality, and producing ICD codes[43,44]. LLMs have also been used for iterative treatment planning optimization. One study demonstrated clinically acceptable Volumetric Modulated Arc Therapy plans for cervical cancer while significantly reducing planning time[45]. In addition to planning, LLMs improved documentation efficiency by automating the generation of CT simulation notes[46]. Lastly, LLMs have been implemented in radiation oncology for question and answering, with research finding that ChatGPT can provide accurate, reliable, and satisfactory answers when responding to radiation oncology specific questions[47,48].

SBRT Stereotactic Body Radiation Therapy (SBRT) is a radiotherapy treatment technique that delivers a high dose of radiation from several directions to a highly conformal target volume surrounding the tumor, while utilizing a steep dose gradient to minimize exposure to the surrounding organs at risk[49,50]. Despite these measures, tumor recurrence and treatment related adverse events (TRAEs) can occur. Radiation pneumonitis and radiation fibrosis within the treatment site and the surrounding healthy tissues[51] are two examples of TRAEs. SBRT is widely used for treatment of early-stage lung cancers. The goal of radiation therapy to maximize the dose to the tumor while not exceeding the tolerance threshold for the organs at risk is complicated due to biological variation between patients, where the two patients with an identical malignancy type will respond differently to the same radiation dose due to factors such as age, genetics, and lifestyle. If an individual requires a dose that exceeds standard safety guidelines to achieve tumor control, adhering strictly to those guidelines may result in underdosing and poor treatment outcomes. Thus, there is a necessity for the development of personalized normal tissue complication probability models. Previous attempts to develop these predictive models have utilized small, site-specific datasets, and fail to accurately capture demographic diversity and the complexity of human biology[52]. Extremely large datasets containing patient data from multiple sites are thus necessary to capture this complexity. LLMs are instrumental to curating these databases. Once established, these databases can be utilized to develop personalized normal tissue complication probability models.

In this study, we leverage LLMs combined with prompt engineering techniques to extract the following clinical endpoints from the radiology notes of patients previously treated with lung SBRT: the original SBRT site, if there is locoregional or distant tumor progression, and radiation induced complications of rib fractures and pulmonary fibrosis. Local progression and radiation related rib fractures represent clinically controllable events occurring within irradiated regions. Both endpoints are rare, and it is therefore necessary to compile and analyze large datasets to identify enough cases for evaluation. While utilizing LLM's for NER and information extraction has been explored in other medical disciplines, they have not been thoroughly investigated in radiation oncology. Our approach is unique in that our target entities are rarely explicitly stated in the radiology report, and the LLM must adhere to specific classification guidelines for accurate classification.

# 2 Methods
## 2.1 Data Acquisition
This retrospective study was conducted under Institutional Review Board (IRB) approval from the University of California, Los Angeles (UCLA) Research Administration. We obtained a dataset containing all primary lung cancer patients who were treated with lung SBRT at UCLA Radiation Oncology clinics between 2008 through 2023. We then applied predefine inclusion and exclusion criteria to create our cohort. Eligible patients were those with primary lung cancer who received lung SBRT. We excluded patients with stage IV disease originating outside the lung who underwent lung SBRT treatment for pulmonary metastases, and patients with a prior history of cancer. We implemented these exclusion guidelines to ensure recorded instances of progression resulted from lung-origin malignancy rather than an additional primary site, which was essential to evaluate our clinical endpoints and the effectiveness of SBRT to treat primary lung cancer.

For prompt development, we randomly selected 30 eligible patients to create a training set consisting of 60 post-SBRT radiology reports. For each patient, two post-treatment reports (chest-CT and /or whole-body PET/CT) were manually extracted from our EHR database into text files stored on our secure institutional network drives to ensure HIPAA compliance. For the test set, we identified an independent cohort of 50 additional eligible patients and similarly extracted two post-SBRT radiology reports (chest CT and/or whole-body PET/CT), producing 100 reports for our test evaluation.

## 2.2 Rulebook Creation and Note Grading

We developed a rulebook detailing our clinical endpoints and decision criteria for our clinical endpoints (Supplementary Figure 1). The radiation-induced toxicities of rib fractures and pulmonary fibrosis were graded as binary outcomes ("Yes" or "No"). Our endpoints of locoregional and distant progression were graded using a three-level scale ("Yes", "No", or "Maybe") to represent indeterminate reporting language and diagnostic uncertainty. Definitions of our clinical endpoints are provided in Supplementary Figure 1.

This rulebook was utilized by the human reviewers to determine the ground truth classifications for our clinical endpoints and model evaluation. The training set was independently reviewed by two separate graders, one American Board of Radiology certified medical physicist and one graduate student researcher. Reviewers first identified the SBRT treated site, which served as anatomic reference for the following progression classifications. Following the criteria defined in the rulebook, the reviewers assessed locoregional and distant progression and determined if the radiology report detailed instances of radiation-induced rib fractures or radiation-induced fibrosis. Once complete, discrepancies were resolved by consensus and the resulting consensus labels were used as the training-set ground truth. The test set was independently reviewed by three graders, and they were blinded to the LLM outputs during the grading process. In addition to the same certified medical physicist and graduate student, a research data analyst within our thoracic radiation oncology program was trained on our rulebook. For the test set, the ground truth was defined as the majority scoring amongst the three independent graders. The LLM's responses were evaluated against the majority-vote.

## 2.3 Prompt Engineering and Analysis

Our goal was to iteratively design and evaluate prompts that instruct the LLM to extract and classify our clinical endpoints from individual radiology reports. We chose the Llama 3.3-70B instruct[53] served through an LLM-specific NVIDIA NIM container, deployed on an Amazon Web Services (AWS) HIPAA compliant P4d.24xlarge instance, consisting of eight NVIDIA A100 40-GB GPUs with 320 GB of total VRAM, 96 vCPUs with 1.1 TB of RAM, and 8 TB of NVMe SSD. Our radiology notes were stored on Amazon S3 and mounted as a filesystem on the running EC-2 instance, allowing our inference pipeline to access our reports and pass them to our NIM-hosted model. We explored modulating the temperature hyperparameter, which is a floating-point value that controls the randomness in token generation by scaling the logits before sampling[54], and the maximum number of output tokens, which determines the length of the LLM's response. A low temperature value close to zero produces highly deterministic outputs by choosing more probable tokens, while higher temperature values produce greater variability in generation. Multiple temperature values were explored, and we selected a near-deterministic temperature value of ≈ 0 as it produced the most accurate and reproducible results on the training set. We also optimized the maximum model output length and set max_tokens to 2048 to ensure the LLM could return a JSON object with supporting evidence and explanations without truncating the response. All model hyperparameters were fixed during inference on the test set.

The LLM input consisted of a system prompt, the radiology report for analysis, and user prompts specifying the tasks. The system prompt defined the LLM's role and behavior by instructing it to act as a radiation oncologist reviewing their patient's thorax radiology report after completion of SBRT. It also defined the LLM's objectives of classifying our clinical endpoints and instructed the model to focus on specific report sections when formulating decisions.

The user prompts contain task-specific instructions derived from our rulebook. For our progression endpoints, the prompts detail standards for locoregional and distant progression, and they include examples of benign post-SBRT findings and other phrasing patterns that the LLM should classify as no-progression. We then employed a few-shot learning component by incorporating example excerpts paired with the ground truth classifications, allowing the model to recognize classification patterns using a small set of example cases. These example excerpts were exclusively sampled from the training set, and the test set was not used for prompt refinement. Additional prompts explained the guidelines for detecting radiation related fibrosis and radiation related rib fractures, along with corresponding labeled examples. Once the classification rules were defined and examples cases provided, the LLM was prompted to analyze the provided radiology report and extract all oncologic findings and clinical findings from the impression section. It was then instructed to determine the original SBRT site within the lung and classify our clinical endpoints while adhering to our benign findings restraints. Finally, the LLM was instructed to return a JSON output for each endpoint, including the predicted classification, quoted evidence to support its decision, and an expressive explanation for its classification.

For each report, the LLM responses were evaluated against the majority reference standard. Identifying the SBRT site was a laterality extraction task, and a classification was scored as correct if the labeled lung matched the human-identified SBRT lung. The LLM's classification of locoregional and distant progression was graded using the confusion matrix categories of true positive (TP), true negative (TN), false positive (FP), and false negative (FN). Since our progression endpoints could be classified as "Yes", "No", or "Maybe", we evaluated two label-mapping schemes. To be conservative, "Maybe" was mapped to "No" for both the reference labels and the LLM outputs to prioritize specificity. Under a sensitive approach to prioritize progression identification, "Maybe" was mapped to "Yes" for both the reference labels and the LLM outputs. Both radiation-induced rib fractures and radiation-induced fibrosis were graded as binary ("Yes" / "No"). For all our clinical endpoints, concordance was graded as either TP or TN, and a mismatch was graded as either FP or FN. These confusion matrix categories were used to calculate sensitivity, specificity, and accuracy. Cohen's kappa (κ) was computed on the binary labels and used to assess inter-observer agreement between the majority gradings and the LLM responses. Fleiss's kappa was employed to determine the inter-observer agreement between the three graders.

We performed a qualitative review of the LLM's classification rationale and the identified excerpts of the radiology note that influenced its response, focusing on the reasoning to understand recurrent failure modes. For correct responses, this approach served as a quality control measure to ensure the model identified the correct report excerpt and applied the correct reasoning outlined in our guidelines. The prompts were iteratively updated after completing this analysis. Our prompt development followed a cycle of updating the prompts, evaluating our prompts on the training set and analyzing the LLM responses (Figure 1). This approach was repeated on our training set until performance no longer increased across our clinical endpoints. The prompts were then frozen for evaluation on our test-set and are shown in Supplementary Figure 2.

.

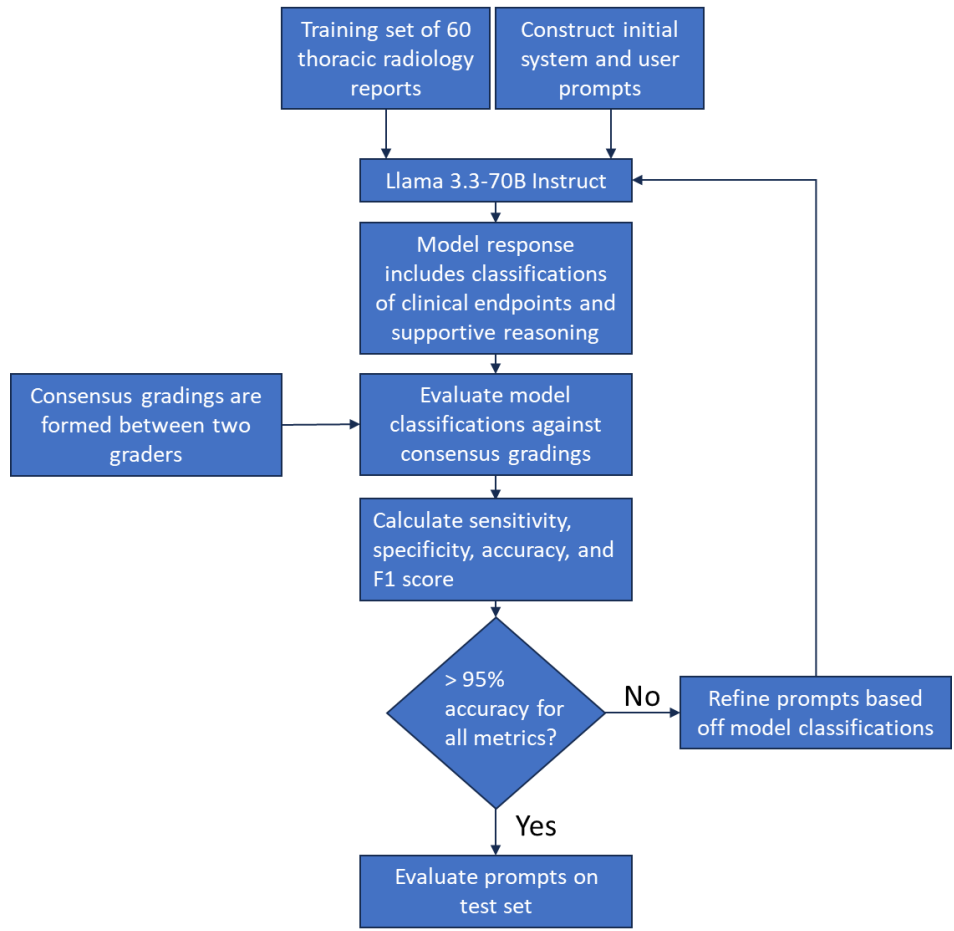

**Figure 1:** Prompt engineering workflow. The LLM prompts are continually refined until > 95% accuracy is achieved across all metrics for all clinical endpoints.

## 3 Results

### 3.1 LLM Classifications

The LLM's overall performance was excellent when identifying radiation-induced TRAEs and is presented in Table 1. The LLM demonstrated a sensitivity of 1, specificity of 0.98, accuracy of 0.98, and κ of 0.80 when identifying radiation-induced rib fractures. For radiation-induced fibrosis, the LLM achieved a sensitivity of 1, a specificity of 0.84, accuracy of 0.90, and a κ of 0.80.

The LLM correctly identified 97% of the sites treated with SBRT. Relative to the majority gradings, the performance of the LLM's clinical endpoint classification is summarized in Table 1, and confusion matrices of the results are shown in Figure 2. For locoregional progression, the LLM performed best when classifications and gradings of "Maybe" were treated as No, achieving a sensitivity 1, specificity of 0.99, accuracy of 0.99, and κ of 0.96. When treating "Maybe" as Yes, performance decreased, with a sensitivity of 0.91, specificity of 0.82, accuracy of 0.84, and κ of 0.61. On the other hand, mapping "Maybe" to Yes improved performance for distant progression. When treating "Maybe" as No, the LLM demonstrated a sensitivity of 0.50, specificity of 0.99, accuracy of 0.97, and κ of 0.70. When "Maybe" was treated as Yes, all metrics improved, as the LLM achieved a sensitivity of 1, specificity of 0.93, accuracy of 0.94, and κ of 0.70.

**Table 1.** LLM responses versus the majority scorings. * Indicates both majority gradings and LLM responses of "Maybe" were graded as "No". † Indicates both majority gradings the LLM responses of "Maybe" were graded as "Yes".

| Clinical Endpoint | Prevalence | Sensitivity | Specificity | Accuracy | κ |
|---|---|---|---|---|---|
| Radiation-Induced Rib Fractures | 0.08 | 1 | 0.98 | 0.98 | 0.88 |
| Radiation-Induced Fibrosis | 0.36 | 1 | 0.84 | 0.90 | 0.80 |
| Locoregional Progression* | 0.13 | 1 | 0.99 | 0.99 | 0.96 |
| Locoregional Progression† | 0.22 | 0.91 | 0.82 | 0.84 | 0.61 |
| Distant Progression* | 0.04 | 0.5 | 0.99 | 0.97 | 0.56 |
| Distant Progression† | 0.08 | 1 | 0.93 | 0.94 | 0.70 |

The three graders achieved substantial agreement across all clinical endpoints, with Fleiss' κ values of 0.70 for locoregional progression, 0.73 for distant progression, 0.84 for radiation-induced rib fractures, and 0.83 for radiation-induced fibrosis.

### 3.2 Model Successes and Failure Modes

To analyze the LLM's behavior beyond our reported metrics, we performed a qualitative analysis of the successes and failures (Table 2). For both locoregional and distant progression, the LLM demonstrated high sensitivity when the radiology reports explicitly described interval changes, new lesions, or expressed concern for progression. Additionally, the model reliably classified no-progression when findings were framed as unchanging post-treatment effects or lesions stable in size. The LLM achieved perfect sensitivity when identifying radiation-related rib fractures and most accurately classified cases when the report included explicit language linking the rib fracture to prior radiation.

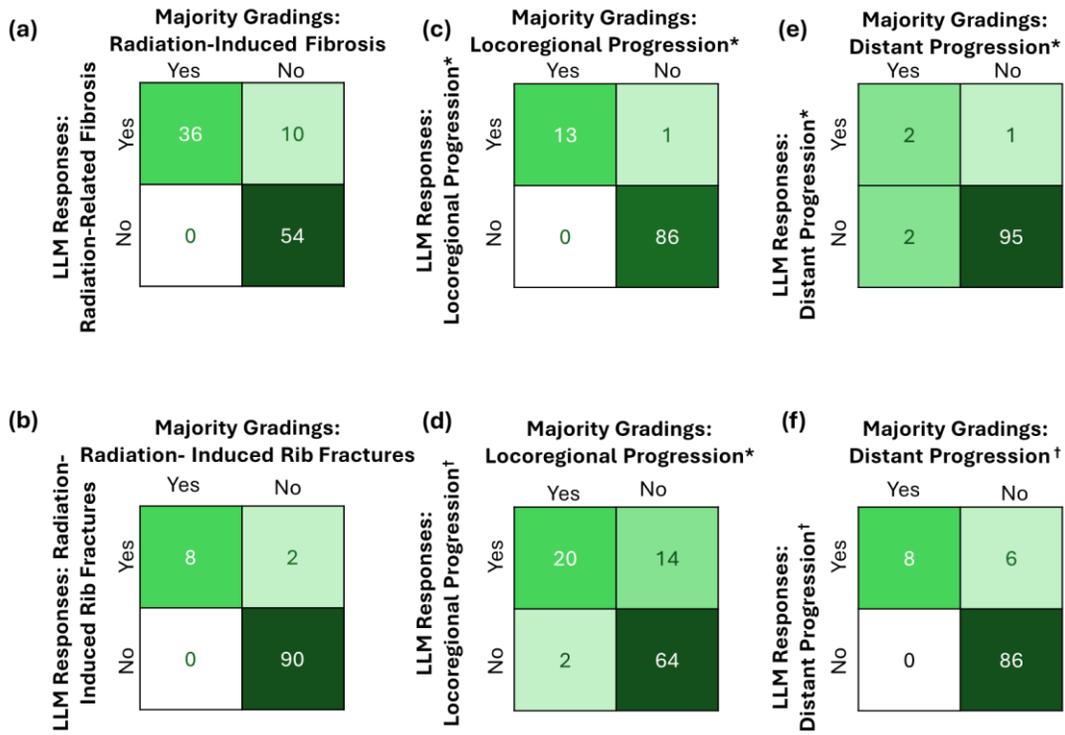

**Figure 2:** Confusion matrices to assess the performance of our LLM to classify our four clinical

Two recurring patterns accounted for most of the progression misclassifications. First, the LLM struggled to distinguish new or enlarging benign post-treatment changes from true disease progression. Second, the model occasionally misinterpreted references to "new" findings that were described as new in a prior report and unchanged in the current study. Errors were more frequent when the radiologist expressed clinical uncertainty, such as ascribing a change to potential treatment effects while suggesting close follow-up for monitoring. In these scenarios, the LLM tended to classify these cases as "Maybe" progression, indicating sensitivity to the uncertain language but failing to strictly adhere to our benign-finding guidelines.

**Table 2.** Examples of LLM Classification for our Clinical Endpoints

| Category | LLM Extracted Radiology Note Excerpt | Majority Grading | LLM Classification |
|---|---|---|---|
| Success: Locoregional | "new costal pleural-based mass/consolidation in the LUL is highly suspicious for progressive malignant disease" | Yes locoregional | Yes locoregional |
| Success: Distant | "Focus of conglomerate nodularity in the left upper lobe with an airway distribution likely representing a focal area of resolving atelectasis" | No distant | No distant |
| Failure: Locoregional | "New ill-defined ground glass ...likely of inflammatory etiology and clinical correlation and interval follow-up is recommended to document resolution." | No locoregional | Yes locoregional |
| Success: Radiation-Induced Rib Fracture | "Interval development of focal deformity and linear lucency of the left anterior second rib that may represent pathologic fracture in the context of radiation osteonecrosis" | Yes radiation-induced rib fracture | Yes radiation-induced rib fracture |
| Success: Radiation-Induced Rib Fracture | "Chronic fractures of the right lateral third and fourth ribs." | No radiation-induced rib fracture | No radiation-induced rib fracture |
| Failure: Radiation-Induced Fibrosis | "Surrounding groundglass opacity and consolidation compatible with radiation changes" | No radiation-induced fibrosis | Yes radiation-induced fibrosis |

## 4 Discussion

### 4.1 Analysis of the LLM Generated Responses

In this work, the LLM excelled at extracting radiation-related toxicities, with near-perfect performance when classifying radiation-induced rib fractures and strong performance identifying radiation-induced fibrosis (Table 1; Figure 2). Radiation-induced rib fractures were often explicitly stated in the radiology note, making them a relatively straightforward extraction task. This is demonstrated by the confusion matrix in Figure 2a, which shows two false positives and zero false negatives. Instances of radiation-induced fibrosis are also frequently mentioned explicitly, and the LLM achieved perfect sensitivity, but produced two false positives (Figure 2b). These failures largely resulted when the model relied on its broader learned concept of fibrosis and linked phrases such as "post-radiation changes" to radiation-induced fibrosis, despite the report not explicitly attributing the finding to fibrosis.

For progression endpoints, the LLM's performance was highest when reports contained direct wording indicating lesion growth, new malignant lesions, or explicit concern for progression. Reports containing benign findings with uncertain language, or language suggesting close follow-up for findings not related to disease progression, degraded performance, as the LLM did not consistently discern these benign findings from true progressive disease. The LLM also occasionally misinterpreted the temporal sequence of clinical findings, for example interpreting "new" as relating the current study when the radiologist intended it to refer to a prior examination. The results suggest that our prompt engineering approach excels at translating clinical findings into structured data when the report contains clear semantic intent, however, the performance of approach decreases when the report contains diagnostic ambiguity or contrasting explanations for clinical finding.

An important result is the treatment of the indeterminant "Maybe" classifications, which significantly impacted the LLM classifications and had different performance impacts on both locoregional and distant progression. This discrepancy indicates a decision threshold that should be determined by the intended use. For locoregional progression, mapping "Maybe" to "No" resulted in the optimal LLM classification performance (Table 1), corresponding to a low false-positive rate while successfully identifying the entirety of the true positives, as reflected in Figure 2 confusion matrices. In contrast, mapping "Maybe" to "Yes" increased the number of false positives (Figure 2), reducing the specificity and overall accuracy, and indicating the LLM tended to classify ambiguous locoregional findings toward progression rather than benign findings.

This trend was reversed for distant progression, where mapping "Maybe" to "Yes" markedly improved the sensitivity (Table 1), indicating the strong ability of the LLM to identify potential or definite instances of distant progression. This reversal is clinically plausible, as the benign post-treatment effects that most often cause radiologists to generate equivocal language most often occur at or near the SBRT treatment site and thus confound locoregional assessments, as distant progression is often unimpacted by these localized post-radiation changes.

The Fleiss' κ values indicate strong, but not perfect, agreement among the human graders across our endpoints. This indicates human disagreement sets the upper limit on the LLM's performance when the reference standard stems from interpretation of the reports. If the radiologists themselves offer ambiguous or hedge upon multiple classifications for a clinical finding, the LLM analyzing that report can be expected to show similar uncertainty.

## 4.2 Justification for LLM Usage

Traditional NLP algorithms, such as Naïve Bayes and Support Vector Machines, are typically optimized for narrowly defined tasks and require extensive feature engineering and retraining to generalize outside of their trained datasets[55–59]. Clinically relevant information in radiology notes may be spread across multiple sections and expressed indirectly or ambiguously, which can degrade rule-based, NER-driven, or traditional machine learning pipelines. Additionally, these approaches often do not provide insight into why a specific classification was made, increasing the difficulty of iterative refinement. Earlier deep learning approaches for natural language processing tasks, such as Recurrent Neural Networks (RNN), Long Short-Term Memory, (LSTM), and Gated Recurrent Unit (GRU) offered improvements over classical machine learning approaches, but were often slow to train, struggled to capture long-range dependences, and suffered performance degradations as document length increased[8,9,60]. These limitations are critical for radiology reports, where import findings may be separated within the note and global understanding is necessary to determine if a finding represents progression. Thus, we selected a locally deployed, pre-trained LLM with prompt engineering to extract our five clinical endpoints. This approach supports long-context reasoning across sections, for example the SBRT site mentioned at the beginning of the note and conclusions related to progression at the end of the note. The LLM's explanation of its classifications enabled efficient iterative prompt engineering and error analysis.

### 4.3 Successful Prompts Engineering Strategies

First, we found that directing the LLM within the system prompt to prioritize specific sections of the radiology report, such as the oncologic findings and the impression sections, improved progression classification performance. The model often overlooked important details within these sections when these instructions were placed in the user prompts. We also observed providing clear, affirmative instructions (what the model should do) was more effective than providing than negative constraints (what the model not do) and helped minimize misclassification errors.

Second, a step-by-step reasoning, chain of thought[61], prompting strategy was particularly effective for our endpoints. This approach encourages the LLM to deconstruct complex problems into intermediate decisions before generating a final answer[61], which improved adherence to the classification constraints and reduced false positives. Furthermore, requiring the LLM to generate brief, detailed explanations for its decisions provided insight into its reasoning process and error patterns, enabling efficient prompt refinement. However, we discovered when the underlying classification was correct, unconstrained responses could become verbose and sometimes drift between competing interpretations, particularly for marginal progression classification cases. For increased consistency, we imposed a structured JSON output with the fixed keys of locoregional and distant progression classifications. This enforcement required an explicit final label for each endpoint, and helped standardize responses, reduced excessively verbose outputs, and discouraged the LLM from fluctuating between classifications.

Lastly, prompt ordering impacted classification accuracy. Our initial prompt structure first defined the clinical endpoints and the model's objectives, followed by the progression guidelines and then constraints. While this approach performed reasonably well, we achieved higher classification accuracy when the specific clinical endpoints were placed at the end of the user prompts. This structure allowed the LLM to first internalize our criteria for progressive disease and benign findings before applying them to our clinical endpoints, resulting in improved classification accuracy.

### 4.4 Limitations and Future Work

This work has several limitations. First, we determined progression status based off a single, randomly chosen report, instead of a set of reports across multiple timepoints. Since radiology impressions account for interval comparisons between notes and evolving trends, analyzing a single note can skew the temporal clinical context and increase ambiguity. Second, the original site treated with SBRT is not reliably detailed in each radiology report following completion of the treatment regimen. This would present a challenge for the LLM, as identifying the lung treated with SBRT is necessary to classify locoregional and distant progression. Similarly, patients with multiple disease sites were excluded for this study, a limitation that may limit generalizability to more real-world datasets. Moreover, the majority scoring formed from the three graders is an imperfect reference standard. While it was the most feasible approach for our study, it contains the uncertainty and bias that is present in the interpretation of the radiology reports.

The prevalence for our progression and radiation-induced rib fracture outcomes was small, leading to class imbalance. This resulted in high accuracy and specificities for both endpoints, as the model performed well on most of the true-negative cases. For progression, the LLM misclassified a subset of the true positive cases. Additionally, our dataset was compiled only with in-house radiology reports and did not contain reports from outside institutions. Differences in report structure between institutions, such as the section headers of oncologic findings and impression, could limit the generalizability of this work. Despite these limitations, this study demonstrates our prompt engineering approach is ready for the automated extraction of radiation-related toxicities from a large database of radiology reports at our institution.

There are multiple avenues for future work. First, we will investigate providing the LLM with a patient's entire set of radiology reports via retrieval augmentation following completion of SBRT to make the progression classification. This may assist in separating treatment related findings from disease recurrence and resolve references to previously new findings. This strategy could allow the LLM to resolve current findings with previous stability and treat findings recommended for follow-up as uncertain rather than definite progression. Second, an ablation study can be performed where different portions of the prompt are removed to determine effectiveness, allowing us to focus on portions of the prompt that impact classification and remove other portions of the prompt that are ineffective. To help with generalization, these prompts can be refined on notes at outside institutions. In the study, we focused solely on using the Llama 3.3 70B-Instruct LLM. There are newer LLMs with improved benchmarks on medical tasks, making these models appealing for our future work.

## 5. Conclusion

Our study demonstrates the high accuracy and feasibility of using LLMs with prompt engineering approaches to extract radiation-induced toxicity data and disease progression classifications from radiology reports.

## Figure S1

### Guidelines for Classifying Progression

**Locoregional progression:** If there is tumor growth at original SBRT site; if radiologist says there is local progression; if radiologist says there is local recurrence or similar phrasing; if there is increased FDG uptake and increase in size of tumor/nodule/node at original SBRT site. Cancer has newly spread in the same lung beyond the original site, to the mediastinal lymph nodes, supraclavicular lymph nodes, thoracic lymph nodes; new nodules or nodules increasing in size in ipsilateral lung, pleura or mediastinal lymph nodes; increased FDG uptake and increase in size of tumor/nodule/node in ipsilateral lung or mediastinal lymph nodes; radiologist reports regional progression, recurrence, similar phrasing

**Distant progression:** new nodules, nodes, tumor masses in the contralateral lung or other parts of the body outside of the lungs; radiologist reports distant progression, recurrence, similar phrasing; increase in FDG uptake and size of tumor/nodule/node in contralateral lung or other parts of the body outside of the lungs

### Guidelines for Classifying Fibrosis and Rib Fracture

**Fibrosis:** report must explicitly state fibrosis or scarring directly related to radiation

**Rib fracture:** report must explicitly state there is rib fracture resulting from radiation or at the site of radiation

### Defining Yes, Maybe, or No When Classifying Progression:

**Phrasing that indicates "Yes":** Compatible with/consistent with/most likely/Suspicious for/probably/likely/concerning for; Recommend tissue sampling

**Phrasing that indicates "Maybe":** Possibly/may represent/moderate increases in size/indeterminate

**Phrasing that indicates "No":** Unlikely/very unlikely/less likely/slight increases in size

### PET/CT Guidelines For Classification of Progression:

**Nodule has increased size:** Yes

**Nodule is stable or decreased in size:** No

**\*\***We are not _directly_ factoring in increases in tracer uptake when classifying progression

**\*\*Note\*\*:** phrasing such as 'slight increase in size' is NOT considered an increase UNLESS the radiologist reports it is concerning for progression, or uses similar phrasing

### Other:

If a nodule, node, or tumor is unchanged in size but is concerning for progression, malignancy, or similar wording, that is NOT considered progression for this note as the tumor did not change in size for this note

# Figure S2

**System Prompt:**

You are a radiation oncologist reading your patient's radiology report following completion of lung SBRT. You have three main goals:

1. Identify the lung treated with SBRT
2. Identify whether there is locoregional progression
3. Identify whether there is distant progression

When identifying each type of progression, you will strictly follow a set of guidelines that provide specific details on the oncologic findings that fit into each progression category. You will also pay strict attention to the set of rules that detail what findings to classify as NO progression.

It is very important to correlate all oncologic findings in the report with the Impression section to further guide your decision-making.

**Prompt 1:**

Guidelines to identify the lung(s) treated with SBRT

Please identify:

- SBRT Lung: (e.g., Right)

**Prompt 2:**

LOCOREGIONAL Progression Guidelines

The following are classified as LOCOREGIONAL progression:

**a.** Lesion/tumor/nodule/mass in the same LUNG as the SBRT site that are EXPLICITLY stated as NEW or INCREASING IN SIZE

- HARD STOP**:** DO NOT INFER THIS. IF YOU ARE UNCLEAR, ASSUME THEY ARE NOT NEW OR INCREASING IN SIZE.

**b.** Disease spread to the following lymph nodes:

- Mediastinal lymph nodes
- Supraclavicular lymph nodes
- Thoracic lymph nodes
- Cardiophrenic lymph nodes

c. Disease spread to the pleura in the SAME LUNG as the SBRT site

d. The radiologist EXPLICITLY mentions local recurrence, local progression, or uses similar wording

e. The radiologist reports regional progression, recurrence, or similar phrasing

**Prompt 3**

DISTANT Progression Guidelines

The following are classified as DISTANT progression:

a. Lesion/tumor/nodule/mass in the OPPOSITE LUNG as the SBRT site that are EXPLICITLY stated as NEW or INCREASING IN SIZE

- HARD STOP: DO NOT INFER THIS. IF YOU ARE UNCLEAR, ASSUME THEY ARE NOT NEW OR INCREASING IN SIZE.

b. Lesion/tumor/nodule/mass outside of the thorax that are EXPLICITLY stated as NEW or INCREASING IN SIZE

- HARD STOP: DO NOT INFER THIS. IF YOU ARE UNCLEAR, ASSUME THEY ARE NOT NEW OR INCREASING IN SIZE.

c. Disease spread to the opposite lung or outside of the thorax

d. The radiologist explicitly states "distant progression" or uses similar wording

**Prompt 4**

BENIGN Findings Guide

1A. Clinical findings that are BENIGN

All of the clinical findings listed below are BENIGN:

a. Treatment/radiation-related effects
b. Post-radiation/radiotherapy changes/response
c. Inflammation of the lung
d. Lesions/nodules/masses with an inflammatory origin
e. Infection OR possible infection in the lung
f. Mass-like opacities
g. Consolidative densities
h. Irregular consolidation patterns
i. Pleural effusions
j. Consolidative opacities
k. Micronodules
l. Fibrosis

1B. Benign findings WITH explicit malignant uncertainty → YES or MAYBE progression

If the benign findings are described WITH explicit language (e.g., "cannot exclude recurrence," "malignant involvement not excluded," "cannot exclude metastatic disease"), they MUST be classified as YES or MAYBE progression.

1C. Benign findings WITHOUT explicit malignant uncertainty → NO progression

If the benign findings are described WITHOUT explicit language (e.g., "cannot exclude recurrence," "malignant involvement not excluded," "cannot exclude metastatic disease"), they MUST be classified as NO progression.

Note: The wording and examples above are preserved from your original text. If you intended 1C to mean "no such explicit malignant-uncertainty phrases are present," I can rewrite 1B/1C to remove ambiguity while keeping your intent.

2. Findings that CANNOT be classified as progression

Clinical findings CANNOT be classified as progression when they are described as:

**a.** Stable
**b.** Unchanged in size
**c.** No significant uptake

**Prompt 5**

Example excerpts from notes and their classifications (LEARN FROM THESE)

**1.** "Right middle lobe mass is grossly unchanged and suspicious for infectious involvement. Multiple additional nodules in the right upper lobe"

- Answer: NO Progression
    - Mass is grossly unchanged and suspected infectious origin
    - Nodules are not explicitly stated as new or increasing in size

**2.** "A right upper lobe mass-like consolidation has interval increased in size. Early recurrence is not excluded, therefore short-term follow-up imaging is recommended"

- Answer: MAYBE Progression
    - Radiologist explains early recurrence is a possibility

**3.** "The growth of the lesion is concerning for malignancy"

- Answer: YES Progression

**Prompt 6**

Progression Classification

Before beginning, please remember that a LINGULAR mass is in the LEFT LUNG.

Let's take this step by step.

1. Qualifier Capture Rule

- When evaluating any finding, you MUST search for and incorporate qualifier language that changes interpretation: "cannot exclude," "not excluded," "concerning for," "suspicious for," or equivalent.
- Evidence quotes must include the qualifier clause. If the qualifier appears in the next or another sentence, include both sentences.
- Do not truncate quotes.

2. Scope/Attachment Rule for Qualifiers

- If the qualifier explicitly names a target (e.g., "right effusion," "left pleura," "this mass"), attach only to that target.
- Else if the qualifier refers to an anatomic region (e.g., "pleural involvement") and multiple findings in that region were mentioned immediately prior, attach the qualifier to ALL compatible findings in that region (including both laterality findings).
- Else default to attaching to the most recent compatible finding.
- When laterality is mentioned earlier (right + left) and qualifier is non-lateralized, represent the qualifier as applying to BOTH unless text explicitly limits it.

3. Extraction Step (DO NOT CLASSIFY YET)

Read through the provided radiology report and extract:

- All oncologic findings, AND
- All findings from the Impression section

DO NOT MAKE ANY PROGRESSION CLASSIFICATIONS in this step.

Additional rule:

- If a sentence lists multiple anatomic sites (comma-separated or joined by "and"), you MUST split it into separate anatomic findings, one per site.

Example:
"FDG avid cardiophrenic, mediastinal, and iliac nodes" becomes:

- FDG avid cardiophrenic lymph nodes
- FDG avid mediastinal lymph nodes

- FDG avid iliac lymph nodes

4. Identify the lung(s) treated with SBRT

Identify the SBRT-treated lung(s).

5. Locoregional Progression Assessment

Using the SBRT lung, analyze the report for locoregional progression strictly following the Locoregional Progression Guidelines. Pay close attention to:

- The Impression section of the report
- ALL benign findings in Benign Findings Guide 1A

6. Distant Progression Assessment

Using the SBRT lung, analyze the report for distant progression strictly following the Distant Progression Guidelines. Pay close attention to:

- The Impression section of the report
- ALL benign findings in Benign Findings Guide 1A

7. Classification Rules

- Locoregional progression = "YES" if ANY anatomic finding meets locoregional criteria
- Distant progression = "YES" if ANY anatomic finding meets distant criteria
- BOTH locoregional and distant may be "YES" in the same report
    - Do NOT choose only one category if evidence supports both
- Classify locoregional progression as YES, NO, or MAYBE
- Classify distant progression as YES, NO, or MAYBE

8. Output Requirement (Valid JSON Only)

{

 "SBRT Site": "",

 "Locoregional Progression": "YES|NO|MAYBE",

 "Distant Progression": "YES|NO|MAYBE"

}

``